\documentclass{article}

\begin{document}
\begin{center}
{\Large\bf On rotating regular nonabelian solutions }
\vspace{1.2cm}\\
J.J.\ van der Bij and Eugen Radu\footnote{\textbf{corresponding author}:
\\Albert-Ludwigs-Universit\"at Freiburg, Fakult\"at f\"ur Physik, 
\\Hermann-Herder-Stra\ss e 3, D-79104 Freiburg, Germany
\\ email: radu@newton.physik.uni-freiburg.de
\\ Telephone: +49-761/203-7630 
\\Fax: +49-761/203-5967}\vspace{0.4cm}\\
\it Albert-Ludwigs-Universit\"at \\
\it Fakult\"at f\"ur Physik\\
\it Freiburg Germany\vspace{0.6cm}\\
\end{center}
\begin{abstract}
A general relation for the total angular momentum of a regular solution of 
the Einstein-Yang-Mills-Higgs equations is derived. 
Two different physical configurations, 
rotating dyons and rotating magnetic dipoles are 
discussed as particular cases. 
The issue of rotating pure  Einstein-Yang-Mills regular solutions
is addressed as well.
Based on the results, 
we conjecture the absence of rotating regular solitons 
with a net magnetic charge.

\end{abstract}

\section{INTRODUCTION}
Extended objects such as non-Abelian monopoles and dyons 
provide a fertile ground for the study 
of the interplay between gravitational and gauge interactions  in the presence
of a spontaneously  broken symmetry. Several results show conclusively that the
gravitational interaction, although weak,  can have significant effects on the
properties of  these objects and therefore, cannot be safely ignored
(for a review and references see \cite{1}).  Most studies assume a spherical
symmetry for the configurations, since  the equations of motion
reduce to coupled ordinary differential equations, that can be solved
relatively easy. An interesting question is whether such solutions can be
generalized to rotating ones. As the equations in this case are partial
differential equations, results are much harder to find. A naive approach is to
use perturbation theory to find slowly rotating solutions. However the absence
or presence of rotating  perturbative solutions, though
indicative, is not in general conclusive to establish the absence or
presence of exact solutions.  For instance in boson star models slowly rotating
perturbative solutions are absent, but solutions with a quantized angular
momentum exist \cite{boson}.  On the other hand in the Einstein-Yang-Mills
(EYM) theory slowly rotating solutions are known to
exist \cite{Brodbeck:1997ek},  but it is not clear
whether they can be extended to exact solutions. It is therefore important to
identify non-perturbative criteria to settle such questions.  

Rotating solutions have been looked for in similar, but slightly different
systems, which should be distinguished. The first are
Yang-Mills-Higgs (YMH) regular solutions, the second black hole
solutions, the third  regular pure EYM solutions.

In the seminal paper of Julia and Zee \cite{Julia-Zee},
a profound  connection 
between the angular momentum and the electric charge 
in a YMH theory has been suggested. 
However, in the absence of gravity, 
it has been shown that Julia-Zee dyons do not
admit  slowly rotating excitations \cite{HSV}.
Nonexistence results have been found also for slowly rotating
regular solutions of EYM theory  with bosonic matter \cite{3}. 
In particular these results apply to the 't Hooft-Polyakov monopole and 
its self-gravitating generalizations.

The situation changes however when one allows for the presence of
an event horizon.
Rotating generalizations of static black hole solutions do exist.
They necessarily have  an electric charge.
In the pure EYM case 
it was shown by Volkov and Straumann \cite{Volkov:1997qb} that, 
within a perturbative approach,
electrically charged 
black holes may exist.
They are   prohibited in the static spherically symmetric case by
no-go theorems \cite{bizon}.
Therefore they are non-static and axially symmetric (see also
\cite{Brodbeck:1997ek}). These rotating black hole solutions
have recently been constructed numerically by Kleihaus and Kunz
\cite{Kleihaus:2001kg}. The configurations possess two independent
parameters: an angular momentum $J$ and an electric charge  $Q$.
They are the EYM analogues of the Kerr-Newmann solution
in Einstein-Maxwell theory.

In the regular case, numerical solutions to the full stationary problem 
are still lacking.
Regular rotating EYM solutions however were obtained 
perturbatively \cite{Brodbeck:1997ek}.
Here the angular momentum and the electric charge were found to be related.
The existence of perturbative rotating solutions came as somewhat 
of a surprise, given the experience with other solitonic solutions. It was
conjectured that this possibility is due to the masslessness of the fields
leading to a powerlike behaviour at infinity. This emphasizes the importance
of the asymptotics  at infinity for the possible existence of exact
solutions.

The aim of this paper is to present a non-perturbative approach to these
problems and to argue that a rotating dyon necessarily contains an event
horizon. This fact appears to be connected with the topological properties of
the configurations (the existence of a net magnetic charge),
as monopole-antimonopole regular configurations with electric charge
and nonzero angular momentum exist. 
Of particular interest are the results in the pure EYM case.
Concerning the rotating case, we shall demonstrate the general validity 
of the relation $J\propto Q$, which was  found within a perturbative approach 
in \cite{Brodbeck:1997ek}.
However, the required asymptotic behavior of the gauge potentials
appears to be incompatible with a nonzero value of the electric 
field at infinity. Thus, we conjecture the absence of rotating regular EYM solutions.

\section{GENERAL FRAMEWORK AND BASIC EQUATIONS}

We follow most of the notations and conventions 
used by Kleihaus and Kunz in their works 
\cite{Kleihaus:2001kg, Kleihaus:1998mn, Hartmann:2001gx}.
The action for a  self-gravitating non-Abelian $SU(2)$ gauge 
field coupled to 
a triplet Higgs field with the usual potential $V(\Phi)$,  is (with G=c=g=1)
\begin{equation} \label{lag0}
S=\int d^{4}x\sqrt{-g}[\frac{\mathcal{R}}{16\pi}-
Tr\{\frac{1}{2}F_{\mu \nu}F^{\mu \nu}+D_{\mu}\Phi D^{\mu}\Phi+
V(\Phi)\}],
\end{equation}
where 
\begin{eqnarray}
D_{\mu}&=&\nabla_{\mu}+i[A_{\mu},\ ], \nonumber
\\
F_{\mu \nu}&=&\partial_{\mu}A_{\nu}-\partial_{\nu}A_{\mu}+i[A_{\mu},A_{\nu} ]. 
\end{eqnarray}
Varying the action (\ref{lag0}) with respect to
$A_{\mu}$, $\Phi$ and $g_{\mu \nu}$, respectively, 
we have the equations 
\begin{eqnarray}
\label{YMeqs}
D_{\mu} F^{\mu \nu} &=& i[\Phi,D^{\nu}\Phi],
\\
\label{Heqs}
D_{\mu}D^{\mu} \Phi &=& i \frac{\delta V}{\delta \Phi},
\\
\label{Eeqs} 
G_{\mu\nu } &=& 8\pi T_{\mu\nu}
\end{eqnarray}
where $G_{\mu\nu}$ is the Einstein tensor and the energy-momentum tensor has the form
\begin{eqnarray} \label{tensor}
T_{\mu\nu} = 2Tr\{F_{\mu \alpha} F_{\nu \beta} g^{\alpha \beta} -\frac{1}{4}
g_{\mu \nu}F_{\alpha \beta}F^{\alpha \beta}
+D_{\mu}\Phi D_{\nu}\Phi 
-\frac{1}{2}g_{\mu \nu}(D_{\alpha}\Phi D^{\alpha}\Phi+
V(\Phi)) \}.
\end{eqnarray}
Since we consider a stationary and axisymmetric system, the spacetime still has 
two Killing vector fields $\xi=\partial/\partial \varphi$ and
$\psi=\partial/\partial t$ 
(also in the asymptotically flat regions of large $r$,
$r$ and $\theta$ are the usual spherical coordinates).   

A metric form satisfying the circularity and Frobenius conditions \cite{Wald-book} reads 
\begin{equation} \label{metric}
ds^2=- f dt^2 +  \frac{m}{f} ( d r^2+ r^2 d \theta^2 )
           +  \frac{l}{f} r^2 \sin ^2 \theta (d\varphi+\frac{\omega}{r} dt)^2,
\end{equation}
where $f,~l,~m$ and $\omega$ are functions on $r$ and $\theta$.
This metric parametrization in terms of isotropic coordinates 
is useful when looking for numerical solutions 
(see $e.g.$ \cite{ boson, Kleihaus:2001kg}).

Asymptotic flatness imposes
on  the metric functions 
the boundary conditions at infinity 
$f= m= l=1, ~\omega=0$.
At the  origin ($r=0$), the boundary conditions on  the metric functions read
$\partial_r f= \partial_r m= \partial_r l= 0, ~\omega= 0$.
For a configuration with parity reflection symmetry,
the derivatives $\partial_\theta f, \partial_\theta m,
\partial_\theta l$ and $\partial_\theta \omega$
 vanish along the $\rho$- and $z$-axis
(with $z=r \cos \theta$ and $\rho = r \sin \theta$).
These boundary conditions apply whatever the matter content of the spacetime.
Note that, as discussed in \cite{Heusler:1996ft}, a metric form (\ref{metric}) 
can be too restrictive for
a general  enough YM ansatz. 
However, in the next two sections, we don't use an 
explicit form of the metric.

The symmetry of the gauge field under a spacetime symmetry means that
the action of an isometry can be compensated by a suitable gauge transformation
\cite{Heusler:1996ft,Forgacs:1980zs}. 

For the Killing vector $\psi$, we choose a gauge such that $\partial A/\partial t$=0.
However, a rotation around the $z-$axis 
can be compensated by a gauge rotation
\begin{eqnarray} \label{Psi}
{\mathcal{L}}_\xi A=D\Psi,
\end{eqnarray}
and therefore
\begin{eqnarray} \label{relations}
F_{\mu \varphi} =& D_{\mu}W,
\\
D_{\varphi}\Phi=&i[W,\Phi] \nonumber,
\end{eqnarray}
where $W=A_{\varphi}-\Psi$.

\section{ANGULAR MOMENTUM IN EYMH THEORY}
The total angular momentum for a regular 
spacetime (\textit{i.e.} no interior boundary)  can be expressed as
\begin{eqnarray} \label{J1}
J&=&\frac{1}{8 \pi}\int R^{\alpha}_{\ \beta} \xi ^{\beta}_{(\varphi)}d^3 \Sigma_{\alpha}
\end{eqnarray}
and, from the  Einstein equations
\begin{eqnarray}\label{J}
J&=&\int T_{\varphi}^{t}\sqrt{-g} d^{3}x
\nonumber\\
&=& \int 2Tr\{F_{r \varphi} F^{r t}
+F_{\theta \varphi} F^{\theta t}+D_{\varphi}\Phi D^{t}\Phi\} \sqrt{-g} d^{3}x
\end{eqnarray}
(equivalently,  $J$ can be read off from the  asymptotic
expansion  of the metric tensor). We show 
that this volume integral can be converted into a surface integral 
in terms of matter fields. 
Using the potential $W$ we find in (\ref{J})
\begin{eqnarray}
T_{\varphi}^{t}\sqrt{-g}&=&2Tr\{ (D_{r}W)F^{rt}\sqrt{-g}
+(D_{\theta}W)F^{\theta t}\sqrt{-g})
+i[W,\Phi] D^{t} \Phi \sqrt{-g} \}
\nonumber\\
&=&2Tr\{D_{r}(WF^{rt}\sqrt{-g})
+D_{\theta}(WF^{\theta t}\sqrt{-g})
\nonumber\\
&^{}&-W(D_{r}F^{rt}+D_{\theta}F^{\theta t})\sqrt{-g}
+i[W,\Phi] D^{t} \Phi \sqrt{-g} \}.
\end{eqnarray}
As a consequence of the YM equations (\ref{YMeqs}) we have also
\begin{eqnarray}
D_{r}F^{rt}+D_{\theta}F^{\theta t}=-D_{\varphi}F^{\varphi t}+i[\Phi, D^t \Phi].
\end{eqnarray}
Also, from the relations (\ref{relations}) we write
\begin{eqnarray}
Tr\{ WD_{\varphi}F^{\varphi t} \}&=&0.
\end{eqnarray}
Making use of the fact that the trace 
of a commutator vanishes
 we obtain 
\begin{eqnarray}\label{T34}
T_{\varphi}^{t}=2Tr\{\frac{1}{\sqrt{-g}}\partial_{\mu}(WF^{\mu t}\sqrt{-g})\}.
\end{eqnarray}
Thus, ignoring possible singularities the expression of the total angular momentum is
\begin{eqnarray}\label{totalJ}
J &=&\oint_{\infty}2Tr\{WF^{\mu t} \} dS_{\mu}
\nonumber\\
&=&-2 \pi \lim_{r \rightarrow \infty}  \int_0^{\pi} d \theta \sin \theta^{~}
 r^2[W^{(r)}F^{(r)}_{rt}+W^{(\theta)}F^{(\theta)}_{rt}+W^{(\varphi)}F^{(\varphi)}_{rt}]. 
\end{eqnarray}
Observe that the above relation is a broader result, since it does not 
depend crucially on the particular form of the symmetry-breaking Higgs potential.
In particular, (\ref{totalJ}) remains valid in the absence of the Higgs field.
Also, this results holds in the flat spacetime limit, following the 
general definition (\ref{J}) of the angular momentum.
Note  that we do not fix the gauge in the derivation of this relation. 

The field equations (\ref{YMeqs}-\ref{Eeqs}) admit classical solutions, 
whose energy is finite and localized in small regions.
Magnetic monopoles and dipoles are example of 
such configurations of gravitating vector and scalar fields,
which can be stable as a result of their nontrivial topology.
The size, the mass and other   features of these configurations
generally have to be  abstracted from the solution of complicated nonlinear
equations.

The extreme nonlinearity of the YMH equations even in flat spacetime
means that they are only integrable in the Bogomol'nyi-Prasad-Sommerfield
(BPS) limit.
The equations in curved spacetime are still more complicated,
and so the only realistic approach is to solve them numerically.
However, in order to evaluate the total angular momentum for a regular
 spacetime we need the asymptotics
of the gauge functions only. 
For asymptotically flat finite energy solutions and \emph{a specific ansatz}
these expressions
can be read off from the field equations.

\section{ROTATING DYONS SOLUTIONS}

A general axially symmetric YMH ansatz has been considered 
for the first time by Manton \cite{7}, and has been generalized 
by Rebbi and Rossi \cite{8} for winding number 
$n>1$ when discussing multimonopole solutions.
The usual ansatz used by 
various authors when discussing axially symmetric 
YMH configurations is derived from the ansatz of Rebbi and Rossi.
In spherical coordinates it reads
 (see also \cite{Hartmann:2000ja}) 
\begin{eqnarray} \label{ansatz-dyons}
A_r=&& \frac{1}{r} H_{1}(r,\theta)\frac{u_\varphi}{2}, \nonumber
\\
A_\theta =&&  (1-H_{2}(r,\theta))\frac{u_\varphi}{2}, \nonumber
\\
A_\varphi =&& -n \sin\theta  \left[H_{3}(r,\theta)\frac{u_r}{2} +
(1-H_{4}(r,\theta))\frac{u_\theta}{2} \right],
\\
A_t =&& \eta \left[H_{5}(r,\theta) \frac{u_r}{2} +
H_{6}(r,\theta) \frac{u_\theta}{2} \right], \nonumber
\\
\Phi =&& \eta \left[\phi_{1}(r,\theta) \frac{u_r}{2} +
\phi_{2}(r,\theta) \frac{u_\theta}{2} \right], \nonumber
\end{eqnarray}
where $\eta$ is the asymptotic value of the Higgs field,
with unit vectors
\begin{eqnarray} \label{vector}
u_{r}      =&& (\sin \theta \cos n\varphi, \ \sin \theta \cos n\varphi, \ \cos \theta) ,
\nonumber\\
u_{\theta} =&& (\cos \theta \cos n\varphi, \ \cos \theta \cos n\varphi, \ -\sin \theta),
\\
u_{\varphi  } =&& (-\sin n\varphi, \ \cos n\varphi, \ 0) \nonumber.
\end{eqnarray}
The winding number $n$ corresponds to the topological 
charge of the solutions \cite{8, 11}.
For $n=1$ the ansatz reproduces the spherically symmetric dyons.  
The ansatz (\ref{ansatz-dyons}) is not the most general one, since it
is obtained by imposing a further discrete  
symmetry $M_{xz} \bigotimes C$, where the first factor represents reflection through the 
$\textit{xz}$-plane and the second factor denotes charge conjugation  \cite{10}.
However, this is the most general dyon ansatz 
fulfilling the circularity condition. 
For this ansatz $\Psi=n \cos \theta \frac{u_{r}}{2} - n \sin \theta \frac{u_{\theta}}{2}$ and
\begin{equation} \label{W-dyons}
W=(-n \cos \theta -n \sin \theta H_3)\frac{u_r}{2}
+n \sin \theta H_4\frac{u_\theta}{2}.
\end{equation}
The boundary conditions at infinity consistent with the requirements
of regularity, finite energy and symmetry, are  \cite{Hartmann:2000ja}
\begin{equation} \label{conditions}
H_{i}=0, \ i=1,2,3,4,6; \ \ \ \
H_{5}=\alpha;\ \ \ \
\Phi_{1}=1;\ \ \ \ \Phi_{2}=0.
\end{equation}
(where $\alpha \leq 1$),
and
\begin{equation} \label{conditions1}
H_{i}=0, \ i=1,3,5,6; \ \ \
H_{i}=1, \ i=2,4; \ \ \
\Phi_{i}=0, \ i=1,2,
\end{equation} 
at the origin.
Given the parity reflection symmetry, we need to consider solutions 
only in the region $0 \leq \theta \leq \pi/2$; on the $z$- and $\rho$-axis
 the functions
$H_1, H_3, H_6, \Phi_2$ and the derivatives $\partial_\theta H_2,\partial_\theta H_4,
\partial_\theta H_5$ and $\partial_\theta \Phi_1$ are to vanish.
\newline
In our units, the magnetic charge of the dyon is $n$.  
Also, when using the electromagnetic  't Hooft field strength tensor
we have the expression for the electric charge \cite{Hartmann:2000ja}
\begin{equation} \label{charge}
Q=\lim_{r \rightarrow \infty} 4\pi \eta r^2\partial_r H_5.
\end{equation}  
The nongravitating axially symmetric dyon solutions of Ref. \cite{Hartmann:2000ja}  
have been found within this ansatz.
As far as we know, their curved spacetime generalization is still missing.

By using the ansatz (\ref{ansatz-dyons}) we find 
\begin{equation} \label{res-dyon} 
 \lim_{r \rightarrow \infty}Tr (r^2 WF_{r t})=-\frac {n  Q}{8 \pi}  \cos \theta 
\end{equation} 
and therefore $J=0$. 
Thus, an axially symmetric regular dyon has a vanishing total angular momentum.

We mention also a different attempt to generate 
(flat-space) axially symmetric YMH solutions 
with a nonzero angular momentum (see \cite{Singleton:1996xc}).
In that approach, the Bogomol'nyi equations are casted in the form 
of the Einstein equations and known exact solutions 
of general relativity are used to generate new YMH solutions.
However, unlike the Kerr solutions, the corresponding 
nonabelian configurations
do not carry any intrinsic angular momentum \cite{Singleton:1996xc}.

However, the dyon solutions do not exhaust the configurations allowed by the
EYMH system. One may ask whether this result remains valid for a different
physical situation, in the absence of a net magnetic charge.   
\section{ROTATING MONOPOLE-ANTIMONOPOLE SOLUTIONS}
As shown by Taubes \cite{taubes}, there are smooth, finite 
action solutions to the SU(2) 
YMH equations in the BPS limit, which do not satisfy 
the first order Bogomol'nyi equations.
A monopole-antimonopole pair (MAP) bound state 
(carrying zero net magnetic charge)  and 
possessing only axial symmetry corresponds to such a 
non-BPS solution \cite{map, Kleihaus:2000sx}.
While dyon solutions reside in topologically nontrivial sectors, 
the MAP pair solution is topologically trivial; in particular it is unstable.

When gravity is coupled to YMH theory, regular MAP solutions 
and black holes with magnetic dipole hair solutions  
have been found numerically \cite{ Kleihaus:2000hx}.
In the flat spacetime limit, the existence of multi-MAP configurations 
with a net electric charge has been predicted in \cite{Hartmann:2000ja}.
The resulting solutions possess magnetic charges of opposite sign, 
but electric charges of equal sign.
It is natural to expect that these flat spacetime solutions can be generalized 
to curved spacetime.

A specific MAP ansatz (consistent also with the circularity condition) 
used in numerical calculations \cite{Kleihaus:2000hx} suplemented 
with a nonvanishing time component of the gauge field reads
\begin{eqnarray} \label{ansatzMAP}
A_r=&& \frac{1}{2r} H_{1}(r,\theta)\frac{u_\varphi}{2}, \nonumber
\\
A_\theta =&&  2(1-H_{2}(r,\theta))\frac{u_\varphi}{2}, \nonumber
\\
A_\varphi =&& -2\sin\theta  \left[H_{3}(r,\theta)\frac{u_r}{2} +
(1-H_{4}(r,\theta))\frac{u_\theta}{2} \right],
\\
A_t =&&  \eta\left[H_{5}(r,\theta) \frac{u_r}{2} +
H_{6}(r,\theta) \frac{u_\theta}{2} \right], \nonumber
\\
\Phi =&& \eta \left[\phi_{1}(r,\theta) \frac{u_r}{2} +
\phi_{2}(r,\theta) \frac{u_\theta}{2} \right], \nonumber
\end{eqnarray}
where, this time
\begin{eqnarray} \label{vector2}
u_{r}      =&& (\sin 2\theta \cos \varphi, \ \sin 2\theta \cos \varphi, \ \cos 2\theta) ,
\nonumber\\
u_{\theta} =&& (\cos 2\theta \cos \varphi, \ \cos 2\theta \cos \varphi, \ -\sin 2\theta),
\\
u_{\varphi  } =&& (-\sin \varphi, \ \cos \varphi, \ 0) \nonumber
\end{eqnarray}
(this ansatz can also be extended to include a winding number $n$ \cite{Kleihaus:2000sx}).

The finite energy, fundamental gravitating MAP solution with a net electric 
charge will satisfy a different set of 
boundary conditions at infinity 
\begin{equation}
H_1=H_2=0, \ \ \
H_3=\sin \theta, \ \ \ 
1-H_4=\cos\theta, \ \ \
H_5=\alpha, \ \ \
H_6=0,  \ \ \
\Phi_1 = 1, \ \ \ 
\Phi_2=0,
\end{equation}
(where $\alpha \leq 1$). We impose at the origin the boundary conditions
\begin{eqnarray}
& H_1=H_3=H_2-1=H_4-1=0\ , &
\nonumber \\
& \sin 2\theta \Phi_1 + \cos 2\theta \Phi_2 =0\ , \ \ \ 
\partial_r
\left(\cos 2\theta \Phi_1 - \sin 2\theta \Phi_2\right) = 0
\ , &
\nonumber\\
& \sin 2\theta H_5 + \cos 2\theta H_6 =0\ , \ \ \ 
\partial_r
\left(\cos 2\theta H_5 - \sin 2\theta H_6\right) = 0. &
\nonumber
\end{eqnarray}
On the $z$-axis the functions 
$H_1, H_3, H_6, \Phi_2$ and the derivatives
$\partial_\theta H_2,\partial_\theta H_4,\partial_\theta H_5, 
\partial_\theta \Phi_1 $ have to vanish,
while on the $\rho$-axis the functions
$H_1, 1-H_4, H_6, \Phi_2$ and the derivatives
$\partial_\theta H_2,\partial_\theta H_3,
\partial_\theta H_5, \partial_\theta \Phi_1$ have to vanish.

For this ansatz $\Psi=\cos 2\theta \frac{u_{r}}{2} - \sin 2\theta \frac{u_{\theta}}{2}$
and 
\begin{equation} \label{W-MAP}
W=(-2 \sin \theta H_3- \cos 2\theta )\frac{u_r}{2}
+(\sin 2 \theta -2 \sin \theta (1-H_4))\frac{u_\theta}{2}.
\end{equation}
From (\ref{totalJ}) we find a relation between the angular momentum and the electric charge 
(defined by (\ref{charge}))
\begin{equation}\label{J-MAP}
J/Q=1.
\end{equation}
This connects  the quantization of charge and angular momentum.
A regular magnetic dipole cannot rotate
unless it is endowed with a net electric charge. 
At this stage we cannot say anything about the magnitude of the magnetic
dipole moment  and the gyromagnetic ratio of the  configuration. This is
clearly an interesting subject but further numerical work is required.

\section{ROTATING EYM REGULAR SOLUTIONS}
An important particular case of the general relation (\ref{J}) 
corresponds to the absence of a  Higgs field.
In the black hole case, Kleihaus and Kunz were recently able to find 
rotating black holes EYM solutions 
within a nonperturbative approach \cite{Kleihaus:2001kg}.
Thus, one may ask whether there are pure EYM globally regular solutions
with a nonzero angular momentum.
As proven above, this fact is related 
to the existence of an electric YM potential.
For the spherically symmetric case, there are no-go theorems 
forbidding these solutions \cite{bizon}.
For the stationary, axisymmetric case, 
some contradictory results have been obtained.

In Ref. \cite{wald}, 
based on some general topological considerations,
it has been conjectured that the Bartnik-McKinnon solutions
are the only stationary
nonsingular solutions of the EYM equations.
However, a perturbative analysis based on the assumption 
of linearization stability has been carried out with a different result.
The authors of Ref. \cite{Brodbeck:1997ek} 
have assumed the existence of a family of stationary regular solutions
of the EYM equations, approaching the static solutions 
for zero angular momentum.
The tangent to this family at $J=0$ satisfies the linearized EYM equations.
The results of their perturbative study imply that
all static spherically symmetric solutions admit
slowly rotating excitations with 
continuous angular momentum $J$ and YM electric charge  
$Q$ proportional to $J$.
Conversely, it is reasonable to expect that for a well-behaved solution
of the linearized equations around the static configurations 
there will be an exact family of rotating solutions.

As far as we know,
no results are available for the regular case,
excepting some preliminary considerations in \cite{Gal'tsov:1998af}. 

A suitable form of the YM connection in this case is
\begin{eqnarray} \label{ansatz-EYM}
A_r=&& \frac{1}{r} H_{1}(r,\theta)\frac{u_\varphi}{2}, \nonumber
\\
A_\theta =&&  (1-H_{2}(r,\theta))\frac{u_\varphi}{2}, \nonumber
\\
A_\varphi =&& -n \sin\theta  \left[H_{3}(r,\theta)\frac{u_r}{2} +
(1-H_{4}(r,\theta))\frac{u_\theta}{2} \right],
\\
A_t =&& \left[H_{5}(r,\theta) \frac{u_r}{2} +
H_{6}(r,\theta) \frac{u_\theta}{2} \right]. \nonumber
\end{eqnarray}
The unit vectors $u_a$ are given by (\ref{vector}); here also
\begin{equation} \label{W-EYM}
W=(-n \cos \theta -n \sin \theta H_3)\frac{u_r}{2}
+n \sin \theta H_4\frac{u_\theta}{2}
\end{equation}
The boundary conditions 
for the magnetic potentials,
familiar from the static configurations \cite{Kleihaus:1998mn}, read
\begin{eqnarray} \label{magnetic-YM-infinity}
H_{1}= H_{3}=0,\ \ \ \
H_{2}= H_{4}=(-1)^{k},
\end{eqnarray}
at infinity, and
\begin{eqnarray} \label{condEYM-origin}
H_{1}= H_{3}=0,\ \ \ \
H_{2}= H_{4}=1,
\end{eqnarray}
at the origin. 
Here $k$ is the node number of the $H_{2}$ and $H_{4}$ functions 
(note the vanishing of the magnetic charge).
The asymptotic expansion of electric potentials is
\begin{eqnarray} \label{electric-YM-infinity}
H_{5} \sim (V -\frac{Q}{r}) \cos \theta, \ \ \
H_{6} \sim(-1)^{k+1}(V -\frac{Q}{r}) \sin \theta.
\end{eqnarray}
Also, for $r=0$ we have
\begin{eqnarray} \label{origin-At}
\partial_r H_{5} = \partial_r H_6=0.
\end{eqnarray} 
The boundary conditions on the $z$-axis are
$H_1 =  H_3 = H_6=0$,
$\partial_\theta H_2 = \partial_\theta H_4 = \partial_\theta H_5=0$,
and agree with
the boundary conditions on the $\rho$-axis,
except for $H_5 = 0$, $\partial_\theta H_6 = 0$.

For the above set of boundary conditions
and the asymptotic expansion (\ref{electric-YM-infinity}),
we obtain from (\ref{totalJ}) the relation
\begin{eqnarray} \label{JQ}
J/Q=4 \pi n.
\end{eqnarray}
Thus, as already obtained within the perturbative approach
\cite{Brodbeck:1997ek},  rotating stationary 
Einstein-Yang-Mills solitons are necessarily electrically charged. 
 
A nonzero asymptotic magnitude of $A_{t}$ is crucial in obtaining this result. 
By combining the existence of an electric potential ($i.e.~F_{\mu t}=D_{\mu}A_{t}$)
 and  using the Yang-Mills equations we find
that the electric term in the  YM energy can be converted 
into a boundary integral and
(see \cite{wald})
\begin{eqnarray} \label{VQ}
VQ=2Tr(\int F_{\mu t} F^{\mu t}\sqrt{-g} d^{3}x).
\end{eqnarray}
Thus a vanishing electric potential at infinity implies no electric field at all.
In the Abelian theory it is possible to gauge $V$ away; in the non-Abelian theory,
such a gauge transformation would render the whole configuration time-dependent.

Stimulated by the results obtained within the  perturbative approach,
we have initiated a numerical study of the problem.
Our methods were similar
to those used by Kleihaus and Kunz in their works \cite{Kleihaus:1998mn, Kleihaus:2001kg}. 
Their scheme solves the  field equations
following an iteration procedure.
One starts with a known static
 configuration and increases
the value of $V$ in small steps.
The field equations are discretized on a 
nonequidistant grid and the resulting system
is solved iteratively until convergence is achieved.
The numerical calculations are performed by using the program
FIDISOL, based on the iterative Newton-Raphson method \cite{FIDISOL}. 
To fix the gauge,  we choose 
the usual gauge condition \cite{Kleihaus:2001kg}
$r \partial_r H_1 - \partial_\theta H_2 = 0$.

The numerical results we found are the following.
First (for $n=1,2,...$), we have found charged, rotating solutions 
for small values of $V$ only,   
typically up to the order $10^{-3}$ 
(the numerical errors increase with the value of $V$ and depends
also on the order of consistency of the differential formulae
for the derivatives).
For higher  $V$, the numerical iteration fails to converge.
Also, there are only small differences between the total energy of these
rotating configurations and those of the corresponding static solutions.
Secondly, we have proven numerically the validity of the relation (\ref{JQ}).

We suspect the source of this behavior to reside 
in the long range behavior of the
magnetic potentials $H_i$. 
For static configurations, the asymptotic solutions
 are familiar from Ref. \cite{Kleihaus:2000ia}.
However, it seems that even a small $V$ will alter these asymptotics. 
For instance, as $r \to \infty$ the equation 
for the magnetic gauge potential
$H_2$ 
\begin{eqnarray} \label{eqH2}
0&=&H_{2, r,r}+\frac{H_{2, \theta, \theta}}{r^2}
-\frac{H_{1, \theta}}{r^2}
+\ln\left(\frac{f \sqrt{l}}{m}\right)_{,r} (H_{2,r}+\frac{H_{1,\theta}}{r})
+\frac{1}{r^2}\ln\left(\frac{f \sqrt{l}}{m}\right)_{,\theta}(H_{2,\theta}-rH_{1,r})
\nonumber\\
&+&\frac{\cot \theta}{r^2}(H_{2, \theta}-rH_{1,r})
-\frac{n^2 m}{l r^2}\left(1-\frac{\omega^2 l \sin^2 \theta}{f^2} \right)
\Big ( H_4 H_{3, \theta}-H_3 H_{4, \theta}
+H_2(H_3^2+H_4^2-1) 
\nonumber\\
&+&\frac{H_2-H_4}{\sin^2 \theta}
+\cot \theta (2H_2 H_3-H_{4,\theta}) \Big )
+\frac{m}{f^2}\Big(H_2(H_6^2+H_5^2)+H_5H_{6, \theta}- H_6 H_{5,\theta} \Big )
\nonumber\\
&-&\frac{\omega m }{r f^2}n \sin \theta
\Big( H_5(H_{4, \theta}-2H_2H_3-\cot \theta (H_2-H_4)
+H_6(H_{3,\theta}-1+2H_2H_4+\cot \theta H_3)
\nonumber\\
&-& \cot \theta (H_{6,\theta}+H_2 H_5)-H_3 H_{6,\theta}-H_4 H_{5,\theta} \Big)
\end{eqnarray}
takes the linearized form
\begin{eqnarray} \label{eqh2}
h_{2, r,r}&+&\frac{h_{2, \theta \theta}}{r^2}
-\frac{H_{1, \theta}}{r^2}-\frac{\cot \theta}{r^2}(rH_{1,r}-h_{2, \theta})
-\frac{n^2}{r^2}\Big((-1)^k H_{3, \theta}+h_2+h_4+(-1)^k 2 \cot \theta H_3
\nonumber\\
&+&\cot ^2 \theta (h_2-h_4)\Big)
+V^2h_2 =0
\end{eqnarray} 
Here $H_2=(-1)^k+h_2$, $H_4=(-1)^k+h_4$; ~$h_2$, $h_4$ 
are small deviations and must decay asymptotically to zero.
For large enough $r$, the last term in (\ref{eqh2}) may not be regarded 
as a "small perturbation".
Thus, it seems that the term $V^2 h_2$ implies oscillatory behavior of the $h_2$ 
rather than an exponential or polynomial decay.
Given these facts, we interpret the numerical results as follows.
For very small $V$ the  $V^2 h_2$ term does not show up in the numerics.
The system therefore mimics the linearized theory. This is consistent with
the quantization of $J/Q$. For larger values of $V$ no convergence is
found anymore, since the solutions do not fall off at infinity.

A similar approch can be applied for the full system.
With the asymptotic values (\ref{magnetic-YM-infinity}, 
\ref{electric-YM-infinity}), 
the reduced system linearizes asymptotically
and we end up with the problem of finding the 
asymptotic behavior of a perturbed linear 
two-dimensional system of equations.
For all equations, the $A_{t}$ components of the gauge 
field act like an isotriplet Higgs field 
with negative metric, and by themselves cause the magnetic components 
to oscillate rather than decrease exponentially as $r \to \infty$.

This fact cannot be obtained  within the perturbative approach 
used in \cite{Brodbeck:1997ek},
since the terms on the form $A_t^2$ are ignored in the first order of perturbation theory.
Thus, if we insist on  the boundary conditions (\ref{magnetic-YM-infinity}),
$V=0$ is required and, from (\ref{VQ}) we obtain a purely 
magnetic nonrotating configuration.
We notice that a related  observation has been made  in
\cite{Kleihaus:2001kg},  where, in the limit of vanishing horizon radius,
no global regular solutions are obtained for $V=0$.

The existence of regular rotating generalizations 
of the Bartnik-McKinnon particles  and their higher 
winding number generalization therefore is highly improbable.
The appearance of oscillating solutions in the perturbative
expansion is a strong indication, that such solutions would not be localized.
For an exact mathematical proof a much more detailed analysis
of the asymptotic solutions would be necessary.
Given the experience with the static 
EYM axially symmetric solutions \cite{Kleihaus:2000ia, Kleihaus:1998mn}, 
this appears to be prohibitively complicated. For instance,
non-analytic terms in $1/r$  have to be considered 
and a separate analysis is required for every $n$.

\section{FURTHER DISCUSSIONS}
The essence of our results 
is a new expression
for the Komar angular momentum in a EYMH theory as a surface integral 
in terms of YM fields.  We have shown that this expression is a useful tool
for the issue of rotating nonabelian solutions.

We point out that many  extensions are possible.
The inclusion of other fields coupled to gauge fields, 
such as a dilaton or a doublet Higgs field is possible 
in the framework of our derivation of (\ref{totalJ}).
In this way, we found that the results 
of  \cite{HSV, 3} remain valid 
in the nonperturbative approach. 
An axially symmetric dyon configuration necessarily
possesses  a vanishing total angular momentum.
However, for a different EYMH configuration,
consisting of a pair monopole-antimonopole
we have predicted that the total angular momentum
is proportional to the electric charge.
These results suggest a connection with the topological properties of the
configurations (the existence of a net magnetic charge).

Of course further refinements in the analysis are possible.
For instance, in the derivation of
(\ref{totalJ}), we assumed regularity of the quantity
$WF^{\mu t}$.
Also,  the ans\"{a}tze we have used are not apriori well defined 
on the $z$- axis and at the origin; 
the question of regularity for the gauge potentials 
is a very intricate one and separate analysis are required for every case
(see $e. g.$ \cite{Kleihaus:2000ia, Kleihaus:1999cc}).

The case of rotating regular EYM solitons is not completely settled,
but we have presented strong
arguments implying the absence 
of rotating regular configurations.
The further investigation of the 
asymptotic behavior of the gauge functions would be desirable.
Furthermore, a study of the EYM system beyond the circular sector of the theory
is  of interest.

When coupled to gravity, the flat spacetime solitons admit also 
"coloured" black hole generalizations.
We can use the above approach to derive the total angular momentum expression
(within a specific YMH ansatz)
 for a rotating dyonic black hole and for a 
 rotating black hole with magnetic dipole hair.
The boundary conditions at infinity are still valid,
 while a new set 
of conditions are to be imposed on the event horizon.
For a rotating dyonic black hole we find this time a nonvanishing $J$, resulting from the 
event horizon contribution. 
Given the inner boundary contribution, the relation (\ref{J-MAP}) is no longer
valid for a rotating black hole with magnetic dipole hair. 
The angular momentum and electric charge are independent quantities. 

The above arguments have been made within a YMH ansatz satisfying the
circularity condition and the proof is therefore not complete for the general
case. However this ansatz is rather general and since the arguments are based
on conservation of angular momentum and the behaviour at infinity of the fields
we expect the results to be valid beyond the specific ansatz within which they
are  derived.
We expect also to obtain a vanishing total angular momentum 
for a regular configuration consisting  of  a different number of 
monopoles and antimonopoles.

We therefore put foreward the following conjecture: \emph{"For any regular
solution in a  gauge theory coupled to gravity, 
a nonvanishing total angular momentum is incompatible 
with a net magnetic charge. Any dyon solution with
a nonzero angular momentum  necessarily contains  an event horizon}."
\\
\\
\\
\newline
{\bf Acknowledgement}
\newline
 This work was performed in the context of the
Graduiertenkolleg of the Deutsche Forschungsgemeinschaft (DFG):
Nichtlineare Differentialgleichungen: Modellierung,Theorie, Numerik, Visualisierung.


\end{document}